\documentstyle[11pt,newpasp,twoside,epsf]{article}
\def\edcomment#1{\iffalse\marginpar{\raggedright\sl#1\/}\else\relax\fi}
\reversemarginpar

\begin{document}
\title{Starbursts and the Evolution of Gas-Rich Galaxies}
 \author{J. S. Gallagher III$^{1}$, C. J. Conselice$^{1}$, L. D. Matthews$^{2}$, N. L. Homeier$^{1}$}
\altaffiltext{1}{Dept. Astronomy, University of Wisconsin, Madison, WI USA}
\altaffiltext{2}{National Radio Astronomy Observatory, 520 Edgemont Rd. 
Charlottesville, VA USA}

\begin{abstract}

Moderately luminous starbursts in the nearby universe often occur in 
disk galaxies that are at most subject to mild 
external perturbations. An investigation of this type of galaxy  
leads to a better understanding of starburst triggering mechanisms 
and the resulting star formation processes, and provides useful 
comparisons to more extreme starbursts seen at high redshifts.

\end{abstract}

\vspace{-1cm}
\section{Introduction}
 
Intermediate scale disk galaxies with starbursts are relatively common in 
the nearby universe. About 5\% of galaxies with $-$17 $<$ M$_{B}$ $< -$21
have evidence for starbursts. Many of these occur in smaller galaxies, 
produce blue colors and strong emission lines, and are in gas rich 
systems, as indicated by their relatively large fractional HI content.

The more luminous nearby blue starbursts structurally 
and spectroscopically resemble the faint blue
compact narrow emission line galaxies (CNELGs), which become common at
redshifts of z$>$0.3 (e.g., Guzman et al. 1998).  
They are less well-matched to high redshift blue galaxies, which 
have a combination of high UV luminosities and huge star formation rates 
that are not common in nearby extreme starbursts.
This difference probably reflects secular evolution in the lives of galaxies; 
for example, systems at high redshift may be subject to frequent, 
strong mergers that trigger hyperactive star formation (e.g. 
Conselice et al. 2001 in prep).

Nearby starbursts frequently show evidence for being dynamically cool,
that is they contain strong spiral arms or bars, features that 
are found in rotationally supported galactic disks. Thus the processes that
produced these starburst did not severely disrupt the disk, or the disk
reformed during the starburst event. Local analogs to the CNELGs include 
profound starbursts arising from minor perturbations (see Figure 1). 

\section{Role of Interactions}

Moderate interactions (glancing collisions between equals, minor mergers) are
apparently the sources of many starbursts. Evidence for this includes 
subtle optical structural features such as wisps, faint
tails, or moderately disturbed kinematics in starbursts, as well 
as the more ubiquitous signatures of disturbed HI. 
Furthermore, in some starbursts where an interaction is the likely trigger, 
the outer regions of the optical disk appear to be relatively symmetric and 
relaxed.   Three possible explanations for this behavior are: 

\noindent $\bullet$ Starburst durations exceed the outer disk relaxation 
times, typically a few rotation periods, or $\geq 0.3$ Gyr.

\noindent $\bullet$ Starbursts occur late during many interactions, perhaps 
as a result of infalling material from tidal tails (e.g., Mihos \& Hernquist 
1996).

\noindent $\bullet$ Perturbations can produce starbursts with minimal impact 
on outer stellar disks; e.g., by stimulating gas inflows from extended 
HI disks. \\



\begin{figure}
\plotfiddle{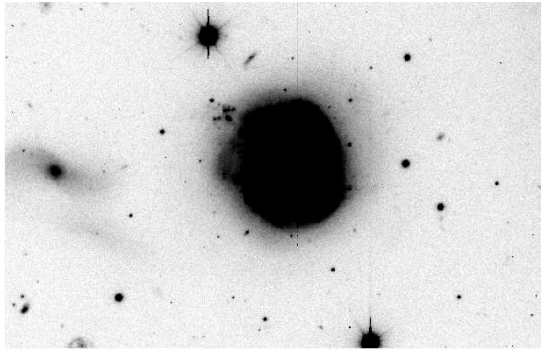}{3.5cm}{0}{33}{33}{-100}{-75}
\caption{WIYN R-band image of the starburst Haro 1 and it distorted 
companion. }
\end{figure}

Our investigations of nearby starbursts lead to several conclusions:

\noindent $\bullet$ Blue starbursts are commonly associated with disk galaxies:
Near face-on systems can transmit UV light through areas where the disk has
been cleaned out by supershells, supernova, etc. (e.g. Conselice et
al. 2000).

\noindent $\bullet$ Low inclination disks can mimic the kinematics of low 
mass galaxies due to
their narrow line widths (Homeier \& Gallagher 1999). Some intermediate
redshift CNELGs could be low inclination disks of moderate mass rather then
extreme dwarf starbursts.

\noindent $\bullet$ Undisturbed intermediate mass galaxies, such as 
extreme late-type and superthin spirals (e.g., Matthews et al. 
1998) can be inefficient star-formers and thereby 
store interstellar gas, the fuel
for starbursts, over cosmic time scales. 

\noindent $\bullet$ UV-bright regions of starbursts frequently occur in 
features associated with disks, such as rings or arms (e.g.
Conselice et al. 2000) which may represent star forming environments not 
yet accessible to severely disturbed high luminosity starbursts 
in the distant universe.

\smallskip

We thank STScI-NASA, the NSF, and the Vilas Trust at the 
University of Wisconsin for support of various aspects of this research.

\smallskip
\leftline{\bf References}

\noindent Conselice, C.J. et al. 2000, AJ, 119, 79 \\
\noindent Guzman, R. et al. 1998, ApJ, 495, L13 \\
\noindent Homeier, N. \& Gallagher, J.S. 1999, ApJ, 522, 199 \\
\noindent Matthews, L., van Driel, W., \& Gallagher, J. 1998. AJ, 116, 1169 \\
\noindent Mihos, C. \& Hernquist, L. 1996, ApJ, 464, 641 \\

\end{document}